# A Programmable Time Alignment Scheme for Detector Signals from the Upgraded muon Spectrometer at the ATLAS Experiment

Jinhong Wang, Liang Guan, J. Chapman, Bing Zhou, Junjie Zhu

*Abstract*—We present a programmable time alignment scheme used in an ASIC for the ATLAS forward muon trigger development. The scheme utilizes regenerated clocks with programmable phases to compensate for the timing offsets introduced by different detector trace lengths. Each ASIC used in the design has 104 input channels with delay compensation circuitry providing steps of ~ 3 ns and a full range of 25 ns for each channel. Detailed implementation of the scheme including majority logic to suppress single-event effects is presented. The scheme is flexible and fully synthesizable. The approach is adaptable to other applications with similar phase shifting requirements. In addition, the design is resource efficient and is suitable for cost-effective digital implementation with a large number of channels.

*Index Terms*—Digital integrated circuits, Clock, Application specific integrated circuits, ATLAS

## I. INTRODUCTION

THE small-strip thin-gap chamber (sTGC) detector will be used for the upgrade of the ATLAS forward muon spectrometer at the Large Hadron Collider (LHC) in 2019 [1]. The sTGC is a gaseous multi-wire proportional chamber and will be used as a trigger and precision tracking device. A simplified structure of the sTGC is shown in Fig. 1, in which a grid of gold-plated tungsten wires are sandwiched between two resistive cathode planes. One cathode plane has strips with a pitch of 3.2 mm × (1-2) m and the cathode plane on the other side is covered with pads of ~8 × 8 cm$^2$. Both pads and strips are used for the muon trigger. The pads provide information about the region of interest while the strips provide precise hit position measurement in the bending plane.

A Trigger Data Serializer ASIC (TDS) has been designed to prepare the trigger data for both pads and strips, perform pad-strip matching, and serialize the data to following circuits [2]. The TDS is a mixed signal ASIC with two modes to handle sTGC pads and strips, denoted

Manuscript received on June. 7$^{th}$, 2017. This work is supported by the Department Of Energy under contracts DESC0008062 and DE-AC02-98CH10886.

The authors are with Department of Physics, University of Michigan, 450 Church Street, Ann Arbor, Michigan, 48109, USA. (e-mail: jinhong@umich.edu).

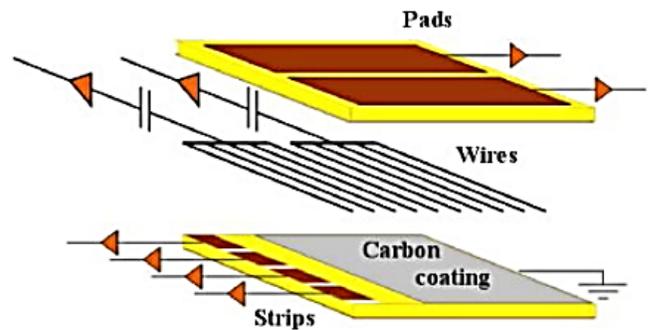

Fig. 1. Structure of the sTGC detector [1].

as pad TDS and strip TDS, respectively. This paper focuses on a specific design issue regarding the pad TDS.

Signals from up to 104 sTGC pads in a quadruplet are routed to two Amplifier-Shaper-Discriminator ASICs (ASD) [3]. The detector signals are converted to Time-Over-Threshold (TOT) pulses and processed by a pad TDS. The pad TDS checks for the occurrence of a TOT pulse in each channel in every LHC beam crossing (BC, ~25 ns cycle) and forwards the Yes/No status of all connected channels to a pad trigger processor. The pad trigger processor uses the pad firing information from multiple sTGC layers to identify a passing muon. Coordinates of the candidate muon tracks are sent to the corresponding strip TDS together with the BC identification (BCID). Deposited charges for strips in the region of interest are then readout, together with the BCID, for muon track reconstruction [2].

The process mentioned above is the first level of muon data suppression at ATLAS in which only a few bands of selected strips are read out from about 350,000 strips. Successful extraction of muon tracks depends on the correct correspondence of pad pulses to their BC. It is thus necessary to eliminate any factors that would distort the timing processing of pad pulses in the pad TDS. One significant factor comes from the delay variations in the routing of the pad signals to a pad TDS, due to the physical size of a pad quadruplet. The delay variation among the 104 pad channels due to different trace lengths is up to 20 ns, and for the proper muon track reconstruction, the variation needs to be compensated to around 3 ns. A delay compensation scheme with a step size of ~ 3 ns in a range

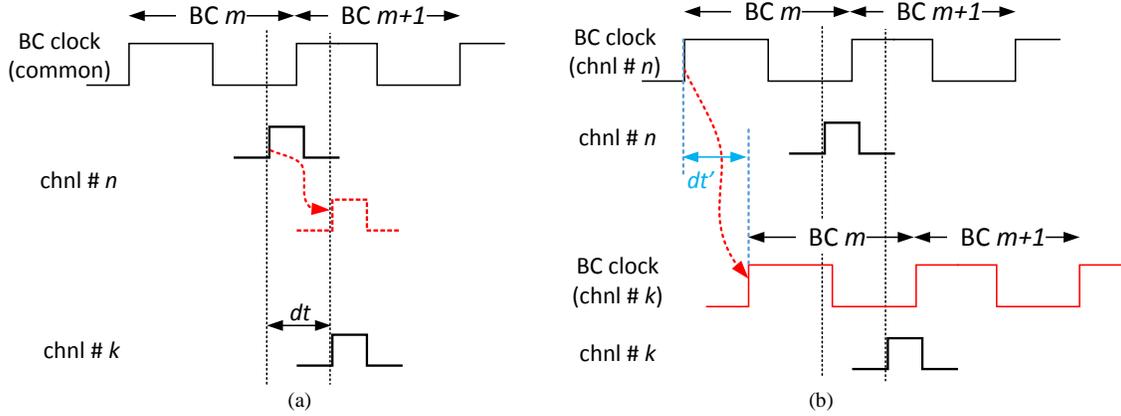
Fig. 2. (a) Delay variation is resolved by adding delay in one channel; (b) Delay variation is compensated by shifting the phase of digitization clock.

of 20 ns for a total of 104 channels is thus essential for the operation of the pad TDS.

A variety of techniques are available for delay compensation. Conventionally, either a delay component from digital cell library or a custom delay circuit with feedback control can be utilized for delay generation [4]. The former relies on sole cell delay and is subject to temperature, supply voltage and process variations, whereas the latter is stabilized with a feedback loop but is complicated in realization and consumes relatively larger silicon area and power, particularly when triple modular redundancy (TMR) is needed to mitigate single-event effect (SEE). These two techniques are not realistic in the TDS implementation due to the large number of inputs (104 channels for pad mode). The requirements on the total power consumption ($< 1$ W), silicon area (cost related), and protection against radiation effects using full TMR all impose challenges for the chip design. Proper choice of the optimal implementation is necessary to meet the stringent specifications.

In this paper, we present a scheme for a resource efficient, flexible, and synthesizable channel delay compensation for the pad TDS. The paper is organized as follows. In Section II, we introduce the techniques used for the synthesizable delay compensation. In Section III, we evaluate the performance in TDS prototypes. We summarize and discuss the general use of the scheme in Section IV.

## II. METHODOLOGY

### A. Implementation Principle

In the pad TDS, the arrival time of a pad pulse is derived from its leading edge. Typically, the timing clock is shared for all channels (BC clock) and delay variations of pulses in two channels are resolved by adding a proper amount of delay in one channel, as shown in Fig.2 (a), where chnl #*n* and #*k* are two channels with an offset of *dt*. An alternative approach used in the TDS is shown in Fig. 2 (b). Individual BC timing clocks are utilized for each pad-TDS channel with programmable phases. The delay variation between channels are compensated by adjusting the relative phases of BC clocks, e.g., separate BC clocks for

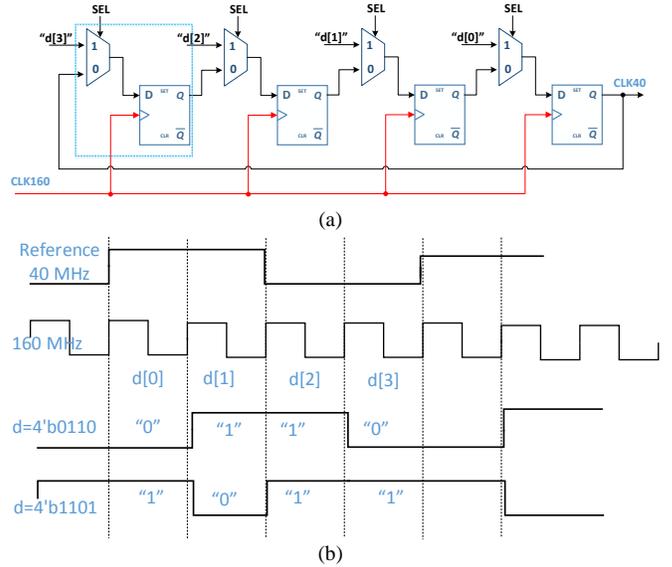
Fig. 3: (a) A four-step phase shift cell; (b) two example phase generation.

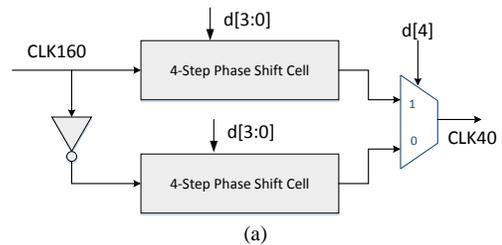

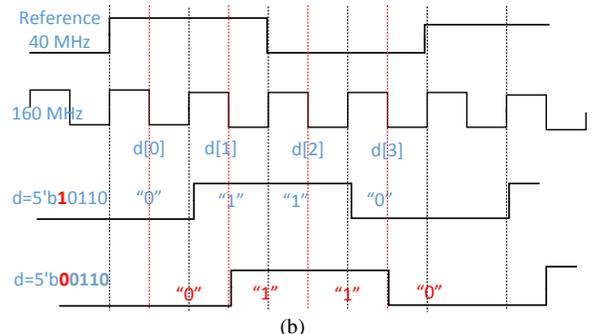
Fig. 4: (a) Interleaving two 4-step cells for 3.125 ns phase shift step; (b) two example phase generation.

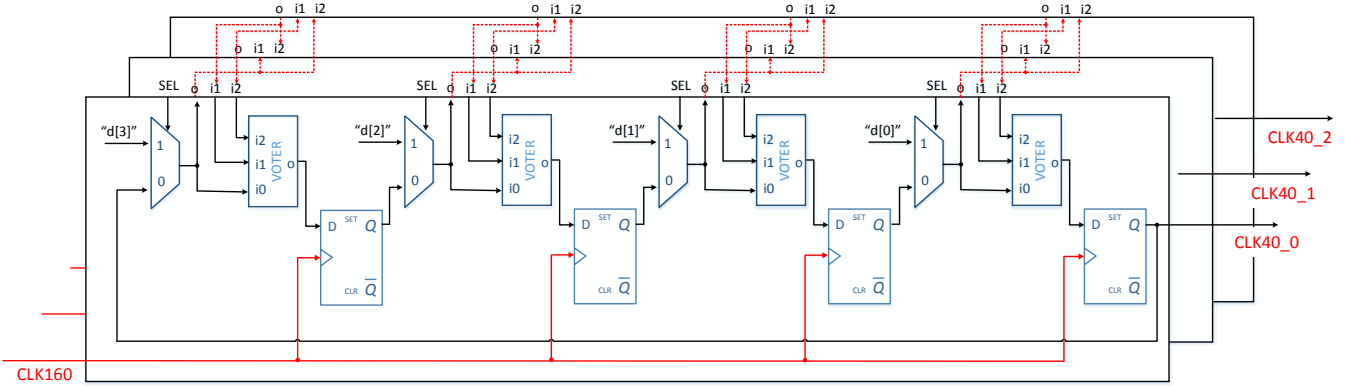

Fig.5. Triple Modular implementation of a 4-step phase shift cell in the pad TDS.

chnl #n and chnl #k are used, and a phase shift of *dt* is applied to chnl #k to compensate for its channel delay. We divide the period of a BC clock (25 ns) into 8 steps to obtain a compensation increment of 3.125 ns.

The 8-step phase shifter is constructed from a 4-step phase shift cell, as shown in Fig. 3 (a). The cell is made up of four shift registers at 160 MHz, the global clock running inside the TDS (CLK160). The phase of the output clock (CLK40) is set by four control bits d[3:0]. Different combinations of d[3:0] result in different phase shifts as well as different high/low ratios. For example, "0110" and "1101" correspond to 6.25 ns and 12.5 ns phase shift respectively with corresponding high/low ratio of 1:1 and 3:1, as shown in Fig. 3(b). The equivalent phase shift step with this cell is 6.25 ns and there are a total of 4 available phases with a 1:1 high/low ratio. For proper operations of following circuits inside the TDS, it is required to keep the high/low ratio of CLK40 at 1:1.

Control bits are loaded either at initialization or at any time during operation via "SEL". It is crucial to load the control bits at a fixed time with respect to the rising edge of the BC reference clock (Reference 40 MHz) to achieve consistent phase shifts for identical control bits. The 4-step phase shift cell depicts the idea of flexible clock phase regeneration. It is synthesizable with a logic resource utilization of only four flip flops in this case.

### B. Phase-shift in the pad TDS

To cope with the 160 MHz global clock in the TDS, the 8-step phase shifter is achieved by interleaving two 4-step shift cells working at the rising and falling edges of the 160 MHz clock respectively. The global clock frequency is determined by the complexity of the TDS logic and the performance of the digital library cells. An illustration of the implementation is shown in Fig. 4 (a), in which two 4-step phase shift cells are utilized and the output clock is selected via another control bit d[4].

An example operation is shown in Fig. 4 (b), in which the generation of two phase shifts $\pi/4$ and $3\pi/8$ are illustrated. Both cells are configured for an output phase $\pi/4$ of CLK40 (d[3:0]="0110"), whereas due to extra phase induced by the inversion of CLK160 ($\pi/8$), an equivalent phase shift of $3\pi/8$ is also obtained. The $\pi/4$ phase is selected via d[4]=1 and $3\pi/8$ is selected with d[4]=0. In the implementation only the high/low ratio of 1:1 is considered.

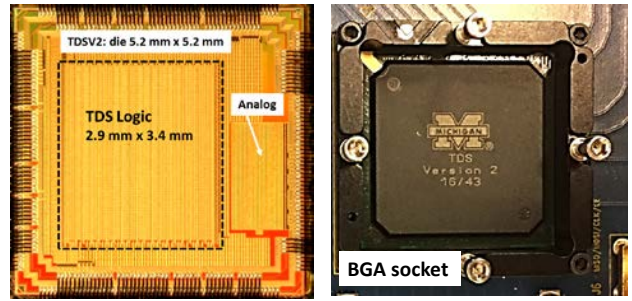

Fig. 6. The TDS die and its package in a BGA socket.

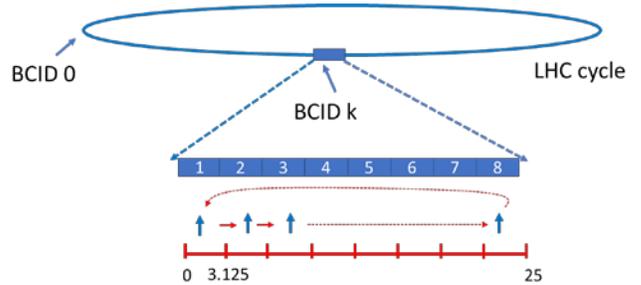

Fig. 7. Illustration of the test bench generation for the pad TDS

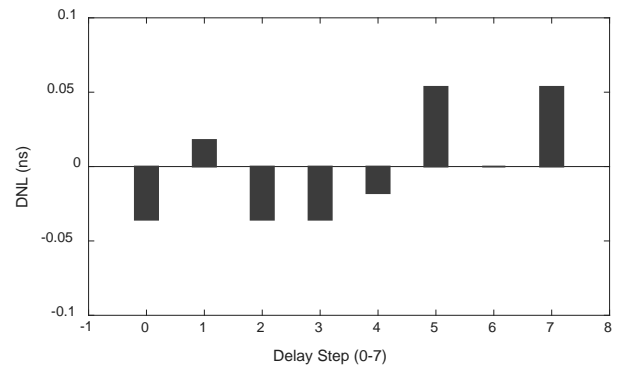

Fig. 9. A typical variation of delay compensation steps in a pad channel (channel 12).

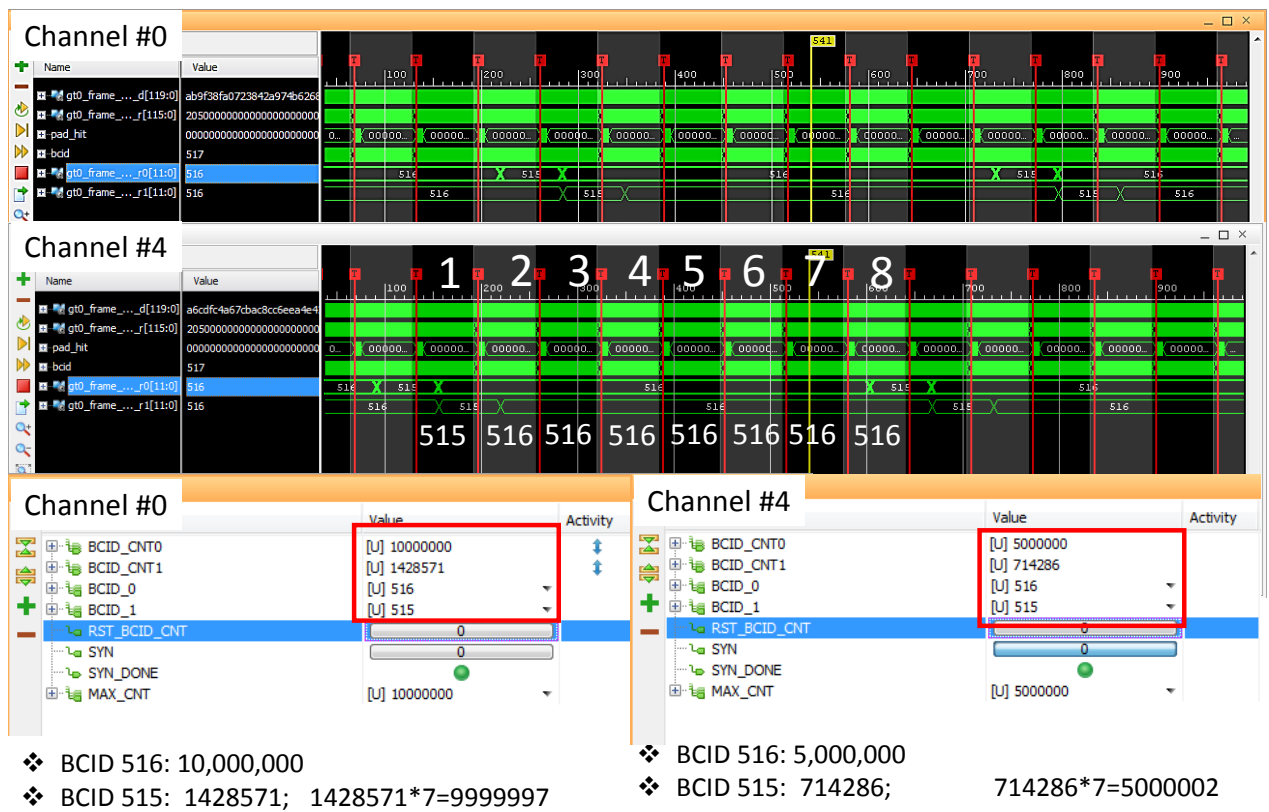

Fig. 8. A typical result from the evaluation of the pad TDS delay compensation.

*C. Circuit Implementation*

The 8-step phase shifter cell is fully synthesizable and the implementation of 104 channels is done by duplicate instantiations in RTL. The logic resource utilization of each phase shifter is 8 flip flops before TMR and 24 flip flops after TMR. In total 2496 sequential registers are utilized for phase shifters in 104 channels, and occupies only 3.25% of the total 76772 sequential registers in the TDS logic.

The TDS will be used on the two innermost stations of the ATLAS endcap muon spectrometer. These two stations are about 7 meters away from the collision point along the beam pipe, and the closest distance to the beam pipe is about 1 meter. The maximum Total Ionizing Dose (TID) expected is about 300 kRad with a safety factor of 6 [5]. This radiation exposure poses no problem for the 130 nm CMOS process that was used [6]. Whereas potentially single-event inducing particles may flip a register or upset a part of logic, disturbing the nominal circuit operation. Full TMR with three voters and triple clock trees are utilized in the TDS to suppress the single event effects, including the phase shifter.

A block diagram of a TMR protected 4-step phase shifter is shown in Fig. 5. The outputs of registers at each stage are voted before feeding into the input of the next stage. The control bits are loaded at initial configuration, and are refreshed every LHC bunch crossing reset (BCR), which circulates at a period of 3580 BC cycles in the current LHC operation. Synchronization circuits are implemented to make sure the refreshing process is performed at a fixed phase with respect to the BC clock and the refreshing process is also free of glitches. The TMR circuit protects the phase shifter from any one bit upset and the refreshing process will bring the circuit back to normal operation in case of multiple bit upsets.

## III. RESULT

We evaluate the performance of the phase alignment scheme in the second prototype of TDS (TDSV2). The silicon area of TDSV2 is about 5.2 mm x 5.2 mm and is packaged in a 400-pin Ball Grid Array (BGA), as shown in Fig. 6. A BGA socket mezzanine board has been designed and fabricated for the performance evaluation. The test is carried out with a Xilnix VC707 FPGA evaluation board. All test benches, control logics and checking of 4.8 Gbps output from TDS are carried out with the VC707 board.

The pad TDS time stamps the inputs of each channel with their corresponding BC clock and serializes the firing status of all 104 channels at 4.8 Gbps every 25 ns. The delay compensation is reflected in the BC time digitization, it is thus not plausible to directly probe any channel outputs nor their BC clocks for evaluation. We utilize a statistical analysis for the performance assessment. It takes 3580 BCs for a beam in the LHC tunnel to traverse a complete LHC cycle. We prepare the test bench by generating a pad pulse only at a particular BC (BCID *k*) for every LHC cycle, as shown in Fig. 7. The pad pulse is stimulated once per LHC cycle and a 3.125 ns delay is induced every time with respect to the previous pulse within BCID $k$. This equivalently divides the BC $k$ into 8 portions with an interval of 3.125 ns, labeled as 1-8 in Fig. 7. A token circulates the 8 slots inside the BC $k$ every LHC cycle and the pulse will be generated from the time slot where the token presents.

The delay evaluation is performed channel by channel. For each channel the test starts from a synchronization between the test bench and TDS, in which the delay compensation step inside the TDS is set to zero while the test bench tunes the phase of the pad pulse generation clock so that all pad pulses captured by the TDC belongs to a single BCID (BCID_0). This is the initial state for the following evaluation. Inside the test bench, two counters (BCID_CNT0 and BCID_CNT1) keep track of the BCIDs from the pad pulses, and BCID_CNT1 counts for the pulses with a BCID other than BCID_0. The initial condition is confirmed once BCID_CNT0 reaches a pre-assigned threshold (MAX_CNT) while BCID_CNT1 remains at 0. The delay evaluation is performed by introducing delay compensation inside the TDS and verifying the ratio of BCID_CNT0 versus BCID_CNT1.

For example, for a 3.125 ns delay compensation, one of the eight consecutive pulses will fall outside of BCID_0, as a result, the corresponding ratio of events with BCID_CNT0 versus BCID_CNT1 is 7, as shown in Fig. 8, in which two example channels (channel #0 and #4) are shown. The ratios are both 7 despite the different MAX_CNT values used. If we denote the eight delay steps inside a pad TDS channel ascendingly as 0-7, for a given step $m$ ($0 \leq m \leq 7$), the corresponding delay ratio of events with BCID_0 and BCID_1 is expected to be $m/(8-m)$. We have performed identical tests for all remaining delay steps and for all 104 channels. The results agree with the expectations.

We further evaluate the exact delay of each compensation step in a channel by tuning the phase of the test bench clock until the added delay inside TDS is completely offset. This is reflected from the ratio turning back to that of the prior delay stage completely. The incremental phase is recorded and it is considered as a close estimation of the size of the current delay step. For the FPGA on the Xilinx VC707 board, the tuning of the clock phase is achieved via the dynamic phase shift of its clock manager circuit [7], with a resolution 1/56 of the VCO frequency (1 GHz in this evaluation, i.e. ~17.86 ps). We repeat the tests for each channel and present a typical result of a channel in Fig. 9. The average delay size is 3.125 ns, with a variation within ~50 ps for all eight steps. The uniformity of the delay step is inherent from the stability of the 160 MHz clock.

IV. CONCLUSION

We present a delay compensation scheme to resolve the time offset among 104 input channels in the pad TDS. The scheme makes use of shift registers to regenerate individual BC timing clocks with programmable phases for each pad channel. The time offset is compensated by a proper choice of the BC timing clock phase for each channel. The scheme is fully synthesizable and logic resource efficient. Protection such as SEE mitigation can be applied via TMR on the HDL source code. The total logic utilization of the design is only about ~3% of the complete TDS logic in terms of registers. The scheme is evaluated in a latest prototype of the TDS with a unique test procedure and the results agree with the design expectations very well.

The delay or clock phase from the scheme is stabilized by the TDS global clock and is thus free of process corners, supply and ambient temperature variations. The unique features provide an optimal solution for the TDS implementation and could also be utilized for other high channel density, cost effective applications.

REFERENCES

[1] ATLAS New Small Wheel Technical Design Report, CERN-LHCC-2013-006, ATLAS-TDR-20-2013, June, 2013.
[2] J. Wang, Trigger Data Serializer ASIC Chip for the ATLAS New Small Wheel sTGC Detector, Nov. 2014, [online] Available: http://cds.cern.ch/record/1969697?ln=en.
[3] G. De Geronimo, J. Fried, S. Li, et al. "VMM1- An ASIC for Micropattern Detectors" *IEEE Trans. On Nuclear Sci.* vol. 60, no.3. pp. 2314-2321, May 2013.
[4] F. Tavernier, S. Bonacini, P. Moreira, "An 8-channel programmable 80/160/320 Mbit/s radiation-hard phase-aligner circuit in 130 nm CMOS", JINST, vol. 7, 2012, c12022.
[5] J. Ameel, D. Amidei, K. Edgar, and L. Guan, "New Small Wheel Radiation and Magnetic Field Environment", available: https://twiki.cern.ch/twiki/pub/Atlas/NSWelectronics/nsw_radiation_bfield.pdf.
[6] F. Faccio, G. Cervelli, "Radiation-Induced Edge Effects in Deep Submicron CMOS Transistors", *IEEE Trans. On Nuclear Sci.* vol. 50, no. 6, pp. *2413*-2420, Dec. 2005.
[7] Xilinx UG472 (v1.13), March 1, 2017